\documentclass[
reprint,
dcolumn,
bm,
nofootinbib,
notitlepage,
superscriptaddress]{revtex4-2}
\usepackage{lmodern}
\usepackage[latin1]{inputenc}
\usepackage[T1]{fontenc}
\usepackage{graphicx}
\usepackage{xcolor}
\usepackage{amsmath}
\usepackage{amssymb}
\usepackage{braket}
\usepackage{mathrsfs}
\usepackage{mathtools}
\usepackage{hyperref}
\usepackage{ulem}
\usepackage{upgreek}
\usepackage{pstricks-add}
\usepackage{natbib} 
\usepackage{listings}
\usepackage{tikz}
\usepackage{tcolorbox}
\usetikzlibrary{mindmap}
\usepackage{cancel}
\usepackage{algorithm2e}
\usepackage{simpler-wick}
\usepackage{pifont}
\newcommand{\xmark}{\ding{55}}%

\definecolor{lightblue}{rgb}{0.145,0.6666,1}

\begin{document}
\title{Spin-cavity interactions in relativistic Jahn-Teller systems under strong light-matter coupling}

\author{Eric W. Fischer}
\email{ericwfischer.sci@posteo.de}
\affiliation{Humboldt-Universit\"at zu Berlin, Institut f\"ur Chemie, Brook-Taylor-Stra\ss e 2, D-12489 Berlin, Germany}

\author{Michael Roemelt}
\email{michael.roemelt@hu-berlin.de}
\affiliation{Humboldt-Universit\"at zu Berlin, Institut f\"ur Chemie, Brook-Taylor-Stra\ss e 2, D-12489 Berlin, Germany}

\date{\today}

\let\newpage\relax

\begin{abstract}
We extend our recent work on the cavity-modified spin Zeeman effect of an effective spin-1/2-system[\textit{J. Chem. Phys.} \textbf{163}, 174307 (2025)] to a relativistic Jahn-Teller scenario under strong light-matter coupling. Here, the effective spin-1/2-system is realized via a single electron or a single hole in a doubly-degenerate molecular orbital system of trigonal symmetric transition metal complexes. Both single-particle and single-hole systems are subject to both vibronic and spin-orbit coupling (SOC) augmented by interactions with a quantized cavity field via the cavity Zeeman interaction. Methodologically, we combine the relativistic $E\times e$-Jahn-Teller model with a recently introduced effective Hamiltonian formalism based on quasi-degenerate perturbation theory, which treats the cavity-spin interaction in leading order beyond the dipole approximation. We derive analytic expressions for Kramers pair energies in weak and strong SOC regimes as well as related cavity-modified effective electronic g-factors. We find cavity-induced modifications of the electronic g-factor to become relevant in the weak SOC regime for both single-particle and single-hole systems while being effectively quenched under strong SOC. Alternating signs of the cavity-Zeeman correction render single-particle and single-hole scenarios distinct in their response to the cavity field from a g-factor perspective. 
\end{abstract}

\let\newpage\relax
\maketitle
\newpage

\section{Introduction}
\label{sec.intro}
The experimental realization of strong light-matter coupling between molecules and confined field modes of Fabry-P\'erot cavities has recently bridged the apparently less related fields of quantum optics and molecular chemistry.\cite{ebbesen2016,garciavidal2021} The flexibility of tuning cavity mode wavelengths to resonantly match molecular excitations allowed for strong coupling involving electronic\cite{hutchison2012}, vibrational\cite{thomas2016,patrahau2024} and spin degrees of freedom\cite{eddins2014,jenkins2016,bonizzoni2017}.   
Strong light-matter coupling with molecules conceptually connects molecular quantum mechanics and (non-relativistic) quantum electrodynamics (QED), which stimulated significant theoretical development of new \textit{ab initio} methods with the aim of rationalizing experimental findings for vibrational and electronic strong coupling regimes from a microscopic perspective.\cite{tokatly2013,ruggenthaler2014,haugland2020,riso2022,schnappinger2023,fischer2024,fischer2025cbocc} 

Spin degrees of freedom and related magnetic properties in the strong coupling regime were in contrast only recently addressed theoretically. Here, one can conceptually distinguish two different approaches as either the Pauli-Fierz Hamiltonian is augmented via effective interactions between spins and classical or quantized magnetic fields \cite{rokaj2019,rokaj2022,barlini2024,fischer2025zeeman}, or the fully relativistic Dirac-Coulomb Hamiltonian is augmented by the non-relativistic quantized cavity field and respective light-matter interactions\cite{konecny2025qedft,konecny2025x2c,thiam2025}. In both cases, a central challenge in the description of cavity-spin-interactions stems from the observation that the commonly employed dipole approximation is no longer valid when magnetic components of the cavity field become relevant.\cite{barlini2024,fischer2025zeeman} We recently addressed this aspect in its simplest form by examining the interaction of a single electronic spin with the quantized cavity magnetic field component subject to a cavity-modified electronic spin Zeeman effect characterized by spin-polariton formation.\cite{fischer2025zeeman} We subsequently discussed cavity-induced modifications of the electronic g-factor in presence of an external classical magnetic field in this ``spin strong coupling'' scenario and elaborated on possible connections to electron paramagnetic resonance (EPR) spectroscopy.

Herein, we extend this initial work to a realistic molecular scenario, specifically, a relativistic $E\times e$-Jahn-Teller (RJT) system\cite{bersuker2006} under strong light-matter coupling. The non-relativistic $E\times e$-Jahn-Teller model characterized by vibronic interactions was only recently investigated in the electronic strong coupling regime.\cite{vendrell2018,nandipati2023} In the relativistic scenario, the Zeeman effect in presence of both classical and cavity magnetic field components is additionally subject to spin-orbit coupling (SOC) effects. Previous studies of the interplay between the Jahn-Teller effect and spin-orbit coupling in Mo(III) and Mo(V) complexes\cite{sharma2017,mcnaughton2010} inspired us to address strong coupling effects on complementary ``single-particle''\cite{sharma2017} and ''single-hole''\cite{mcnaughton2010} systems with $^2E$ electronic ground states. Methodologically, we exploit the velocity-gauge perspective and an effective Pauli-Fierz Hamiltonian formalism in combination with quasi-degenerate perturbation theory\cite{mcweeny1965,neese1998,neese2017}. This approach allows us to obtain cavity-induced corrections of the electronic g-factor for single-particle and single-hole systems, which we discuss in limiting scenarios of weak and strong SOC. 

The remainder of this work is structured as follows. In Sec.\ref{sec.theory}, we introduce relativistic $E\times e$-Jahn-Teller models for single-particle/-hole systems and discuss their basic properties with a focus on the electronic g-factor. We subsequently present an extension of the RJT model accounting for the cavity Zeeman interaction in Sec.\ref{eq.cRJT_model} and discuss relevant basis states and their mutual interactions. We then combine exact diagonalization and QDPT techniques in Sec.\ref{sec.cavity_modified_rjt} to derive cavity-induced modifications of the electronic g-factor and discuss their properties relative to the bare molecular scenario. Finally, Sec.\ref{sec.conclusion} concludes this study.

\section{Relativistic $E\times e$-Jahn-Teller effect}
\label{sec.theory}

\subsection{Molecular Model Systems}
\begin{figure}[hbt!]
\includegraphics[scale=1.0]{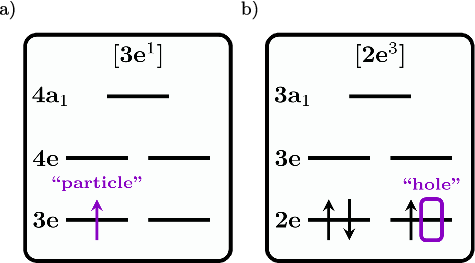}
\caption{Frontier d-orbitals of molybdenum (Mo) in a trigonal symmetric $(C_{3v})$ transition metal complex for (a) ``single-particle'' $3e^1$-configuration of Mo(V) studied in Ref.\cite{sharma2017} and (b) ``single-hole'' low-spin $2e^3$-configuration of Mo(III) studied in Ref.\cite{mcnaughton2010} with $3e=[d_{xz},d_{yz}]$, $4e=[d_{xy},d_{x^2-y^2}]$ and $4a_1=[d_{z^2}]$ (same energetic ordering of $2e$, $3e$ and $3a_1$ orbitals).}
\label{fig.configs}
\end{figure}

We consider the relativistic $E\times e$-Jahn-Teller effect in trigonal symmetric transition metal complexes. Specifically, we concentrate on two complementary scenarios subject to a doubly-degenerate $^2E$ electronic ground state as realized in [Mo]NH with a $d^1$-configuration (Mo(V))\cite{sharma2017} and in [Mo]N$_2$ with a low-spin $d^3$-configuration (Mo(III))\cite{mcnaughton2010}. In Figs.\ref{fig.configs}a and b, we illustrate Mo frontier d-orbitals realizing related single-particle and single-hole scenarios, $[3e^1]$ and $[2e^3]$, respectively. 

In both cases, the ground state is determined by a degenerate orbital doublet, $e=[d_{xz},d_{yz}]$, with z-component of the total orbital angular momentum, $m_l=\pm 1$, subject to a Jahn-Teller distortion induced by vibronic coupling to doubly degenerate vibrational modes. Spin-orbit coupling manifests distinctly in both system via antiferromagnetic interactions for the single-particle case and ferromagnetic interaction for the single-hole case as outlined in the following Sec.\ref{sec.rjt_hamilton}.  
\begin{figure*}[hbt!]
\includegraphics[scale=1.0]{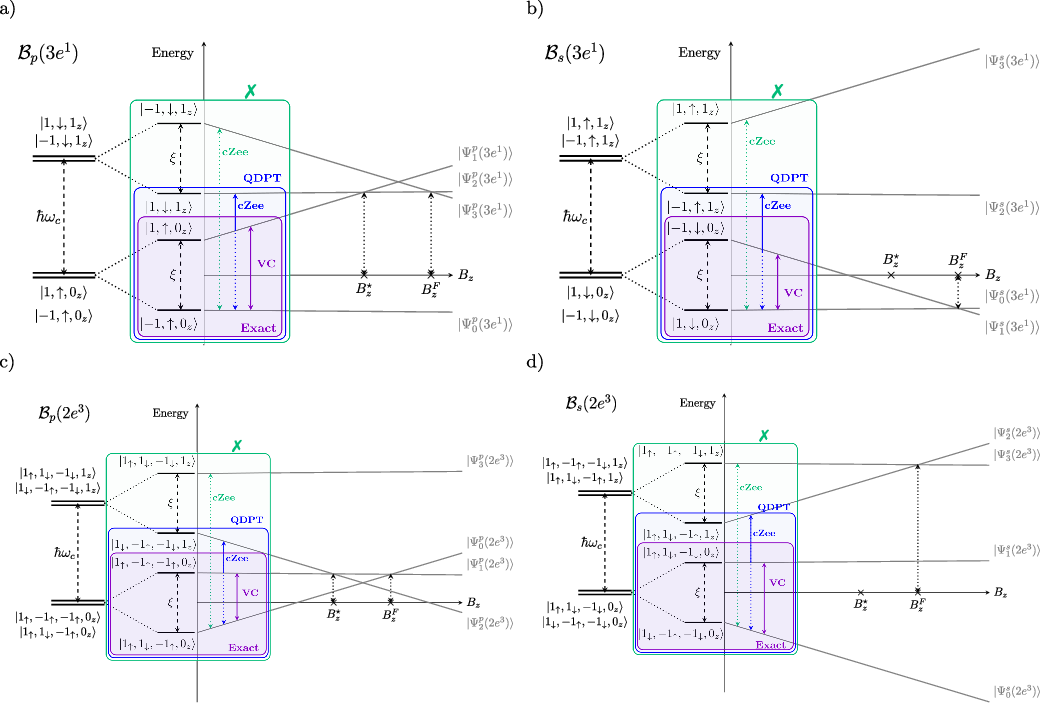}
\caption{Energy schemes for polariton and spectator basis states for (a,b) single-particle ($3e^1$) and (c,d) single-hole ($2e^3)$ scenarios and their dispersion with respect to classical B-field amplitude. Coloured boxes illustrate solution strategy for respective lowest-lying state in weak $B_z$-field limit with vibronic coupling (VC) in lowest-lying two-dimensional subspace being treated exactly (``Exact'', purple box), indirect cavity Zeeman interaction (cZee) of second excited state with ground state captured via quasi-degenerate perturbation theory (QDPT, blue box) and energetically highest-lying state being neglected due to energetic reason (\xmark, green box).}
\label{fig.crjt_polariton_spectator_dispersion}
\end{figure*}

\subsection{Relativistic Jahn-Teller Hamiltonian}
\label{sec.rjt_hamilton}
The RJT model Hamiltonian is given by
\begin{align}
\hat{H}_\mathrm{RJT}
&=
\hat{H}_\mathrm{soc}
+
\hat{H}_\mathrm{Zee}
+
\hat{H}_\mathrm{vc}
\quad,
\label{eq.rjt_hamilton}
\end{align}
with an effective SOC contribution, $\hat{H}_\mathrm{soc}$, the classical Zeeman interaction involving orbital and spin angular momentum, $\hat{H}_\mathrm{Zee}$, as well as the vibronic coupling (VC) term, $\hat{H}_\mathrm{vc}$. A detailed discussion of operators is provided in Appendix \ref{sec.details_rjt_hamilton}, while we will concentrate here on suitable basis sets modelling single-particle and single-hole scenarios followed by a discussion of the related matrix representation of Eq.\eqref{eq.rjt_hamilton}. The $[3e^1]$ configuration of the Mo(V) ion can be described by a four-dimensional product basis
\begin{align}
\mathcal{B}_\mathrm{RJT}(3e^1)
=
\{\ket{1,\uparrow},\ket{-1,\uparrow},\ket{1,\downarrow},\ket{-1,\downarrow}\}
\quad,
\label{eq.rjt_basis_e1}
\end{align}
composed of orbital angular momentum and spin angular momentum eigenstates (z-projection), $\ket{\pm 1}$ and $\ket{\uparrow},\ket{\downarrow}$, respectively. For the $[3e^3]$ configuration realized in the Mo(III) ion, we consider a slightly more complex four-dimensional product basis
\begin{multline}
\mathcal{B}_\mathrm{RJT}(2e^3)
=
\left\lbrace
\ket{1_\uparrow,1_\downarrow,-1_\uparrow},
\ket{1_\uparrow,1_\downarrow,-1_\downarrow},
\right.
\\
\left.
\ket{1_\uparrow,-1_\uparrow,-1_\downarrow},
\ket{1_\downarrow,-1_\uparrow,-1_\downarrow}
\right\rbrace
\quad,
\label{eq.rjt_basis_e3}
\end{multline}
with the abbreviation, $\ket{1_\uparrow}=\ket{1,\uparrow}$. In both scenarios, the RJT Hamiltonian is block-diagonal in spin space with matrix representation (\textit{cf.} Appendix \ref{sec.details_rjt_hamilton})
\begin{align}
\underline{\underline{H}}_\mathrm{RJT}
&=
\begin{pmatrix}
\underline{\underline{H}}_\uparrow & \underline{\underline{0}} \\
\underline{\underline{0}}          & \underline{\underline{H}}_\downarrow
\end{pmatrix}
\quad,
\label{eq.rjt_hamilton_basis}
\end{align}
where the $\uparrow$-sector is described by
\begin{align}
\underline{\underline{H}}_\uparrow
=
\begin{pmatrix}
-
\frac{\xi }{2}
\mp
(1\mp\frac{g_e}{2})\mu_B B_z
& 
F\rho
\vspace{0.2cm}
\\
F\rho 
& 
\frac{\xi}{2}
\pm
(1\pm\frac{g_e}{2})\mu_B B_z
\end{pmatrix}
,
\label{eq.rjt_hamilton_basis_up}
\end{align}
while the $\downarrow$-sector relates to
\begin{align}
\underline{\underline{H}}_\downarrow
=
\begin{pmatrix}
-
\frac{\xi}{2}
\pm
(1\mp\frac{g_e}{2})\mu_B B_z
& 
F\rho 
\vspace{0.2cm}
\\
F\rho 
& 
\frac{\xi}{2}
\mp
(1\pm\frac{g_e}{2})\mu_B B_z
\end{pmatrix}
.
\label{eq.rjt_hamilton_basis_down}
\end{align}
On the diagonal, both expressions are determined by an effective SOC strength, $\xi$, the free electronic g-factor, $g_e$, the Bohr Magneton, $\mu_B$, and the classical B-field amplitude, $B_z$, for an external B-field aligned parallel to the $z$-axis. Off-diagonal matrix elements reflect vibronic coupling between states differing in orbital angular momentum with interaction strength, $F$, and polar normal mode coordinate, $\rho$, where we exploit rotational symmetry of the linear JT model to neglect the azimuthal component, $\phi$. Notably, we combine here results of the two basis sets in Eqs.\eqref{eq.rjt_basis_e1} and \eqref{eq.rjt_basis_e3} where upper signs refer to the single-particle scenario while lower signs reflect the single-hole case differing only in the classical Zeeman interaction. 

Lower-lying eigenstates of distinct spin-sectors with energies, $E^\uparrow_0$ and $E^\downarrow_0$, determine the Zeeman splitting in the RJT model  
\begin{align}
\Delta_\mathrm{Zee}
&=
E^\uparrow_0-E^\downarrow_0
=
g_\mathrm{eff}
\mu_B
B_z
+
\mathcal{O}(B^3_z)
\quad,
\label{eq.rjt_zeeman_split}
\end{align}
in the weak $B_z$-field limit (\textit{cf.} Appendix \ref{sec.details_rjt}). The effective electronic g-factor is obtained as
\begin{align}
g_\mathrm{eff}
&=
\left.
\dfrac{1}{\mu_B}
\dfrac{\partial \Delta_\mathrm{Zee}}{\partial B_z}
\right\vert_{B_z=0}
=
g_e
\mp
\dfrac{2\xi}{\sqrt{\xi^2+4 F^2\rho^2}}
\quad,
\label{eq.rjt_gfac}
\end{align}
where the negative correction corresponds to the single-particle scenario and the positive correction to the single-hole counterpart. We shall now specifically address two limiting regimes determined by effective SOC and VC strengths, $\xi$ and $F$, with 
\begin{align}
g_\mathrm{eff}
&=
g_e
\mp
\dfrac{\xi}{F \rho}
+
\mathrm{O}(\xi^2)
&,\quad
\xi
\ll
F\rho
\,,
\label{eq.geff_weakSOC}
\vspace{0.2cm}
\\
g_\mathrm{eff}
&=
g_e
\mp
2
\pm
4
\dfrac{F^2\rho^2}{\xi^2}
+
\mathrm{O}(F^3)
&,\quad
\xi
\gg
F\rho
\,.
\label{eq.geff_strongSOC}
\end{align}
In the weak SOC regime $(\xi\ll F\rho)$, the electronic g-factor is only slightly modified by a correction linear in the SOC strength, $\xi$, for both electron configurations. In contrast, the strong SOC regime $(\xi\gg F\rho)$ is characterized by antiferromagnetic and ferromagnetic coupling between spin and orbital angular momentum for the single-particle and single-hole scenarios, respectively. In the former case, the free electronic g-factor is nearly quenched opposed to an enhancement under ferromagnetic coupling. The leading-order correction in the vibronic coupling strength, $F$, is therefore especially relevant for the antiferromagnetic scenario.

\section{Cavity RJT Hamiltonian}
\label{eq.cRJT_model}
\subsection{Cavity Zeeman Interaction in RJT Systems}
\label{sec.rjt_cavity_zeeman}
We are now in the position to extend the RJT model Hamiltonian in Eq.\eqref{eq.rjt_hamilton} to the strong coupling regime by accounting for the quantized cavity field and the corresponding Zeeman interaction. Motivated by the observation that the RJT model Hamiltonian is diagonal in $z$-component spin space and recent findings on the relevance of polarization for cavity-spin interactions\cite{fischer2025zeeman}, we consider an effective single cavity mode polarized along the $z$-axis. The corresponding augmented model Hamiltonian reads
\begin{align}
\hat{H}_\mathrm{cRJT}
&=
\hat{H}_\mathrm{RJT}
+
\hat{H}_c
+
\hat{H}_\mathrm{cZee}
\quad,
\label{eq.cavity_rjt_hamilton}
\end{align}
with zero-point energy shifted effective single-mode Hamiltonian
\begin{align}
\hat{H}_c
&=
\hbar\omega_c
\hat{b}^\dagger_z
\hat{b}_z
\quad,
\label{eq.cavity_hamilton}
\end{align}
determined by the harmonic cavity frequency, $\omega_c$, besides photonic creation and annihilation operators, $\hat{b}^\dagger_z$ and $\hat{b}_z$, respectively. The cavity Zeeman interaction, $\hat{H}_\mathrm{cZee}$, for a $z$-polarized cavity mode was recently derived as\cite{fischer2025zeeman}
\begin{align}
\hat{H}_\mathrm{cZee}
&=
\mathrm{i}
\dfrac{g_0 g_e\mu_B}{2c}
\sqrt{\dfrac{\hbar\omega_c}{2}}
\sigma_y
\left(
\hat{b}^\dagger_z
-
\hat{b}_z
\right)
\quad,
\label{eq.cavity_zeeman_op}
\end{align}
with Pauli-Y-operator, $\sigma_y=\mathrm{i}(\ket{\downarrow}\bra{\uparrow}-\ket{\uparrow}\bra{\downarrow})$, speed of light, $c$, and light-matter interaction strength, $g_0$, respectively. The cavity Zeeman interaction couples in this scenario distinct spin sectors via excitation of a cavity photon, which can induce the formation of spin-polariton states.\cite{fischer2025zeeman} Moreover, Eq.\eqref{eq.cavity_zeeman_op} results from a first-order correction augmenting the commonly employed dipole approximation.\cite{barlini2024,fischer2025zeeman}

\subsection{Polariton and Spectator Problems}
\label{sec.rjt_pol_spec}
We augment the RJT basis in Eqs.\eqref{eq.rjt_basis_e1} and \eqref{eq.rjt_basis_e3} now by zero- and one-photon states, $\ket{0_z}$ and $\ket{1_z}$, as
\begin{align}
\mathcal{B}_\mathrm{cRJT}(X)
=
\mathcal{B}_\mathrm{RJT}(X)
\otimes
\{\ket{0_z},\ket{1_z}\}
\quad,
\label{eq.cavity_rjt_basis}
\end{align}
with configurations, $X=3e^1,2e^3$, which ultimately resembles an eight-dimensional model space for single-particle and single-hole scenarios. Inspection of the different interactions in $\hat{H}_\mathrm{cRJT}$ allows us to decompose the eight-dimensional basis, $\mathcal{B}_\mathrm{cRJT}$, into two four-dimensional subsets 
\begin{align}
\mathcal{B}_\mathrm{cRJT}(X)
&=
\mathcal{B}_p(X)
\oplus
\mathcal{B}_s(X)
\quad,
\label{eq.crjt_bibasis}
\end{align}
with a polariton component
\begin{align}
\mathcal{B}_p(3e^1)
&=
\begin{Bmatrix}
\ket{-1,\uparrow,0_z} & \overset{\mathrm{VC}}{\Leftrightarrow} & \ket{1,\uparrow,0_z}
\vspace{0.2cm}
\\
\Updownarrow \mathrm{cZee} & & \Updownarrow \mathrm{cZee}
\vspace{0.2cm}
\\
\ket{-1,\downarrow,1_z} & \overset{\mathrm{VC}}{\Leftrightarrow} & \ket{1,\downarrow,1_z}
\end{Bmatrix}
\label{eq.polariton_basis_e1}
\vspace{0.2cm}
\\
\mathcal{B}_p(2e^3)
&=
\small{
\begin{Bmatrix}
\ket{1_\uparrow,1_\downarrow,-1_\uparrow,0_z} & \overset{\mathrm{VC}}{\Leftrightarrow} & \ket{1_\uparrow,-1_\uparrow,-1_\downarrow,0_z}
\vspace{0.2cm}
\\
\Updownarrow \mathrm{cZee} & & \Updownarrow \mathrm{cZee}
\vspace{0.2cm}
\\
\ket{1_\uparrow,1_\downarrow,-1_\downarrow,1_z} & \overset{\mathrm{VC}}{\Leftrightarrow} & \ket{1_\downarrow,-1_\uparrow,-1_\downarrow,1_z}
\end{Bmatrix}
}\normalsize
\label{eq.polariton_basis_e3}
\end{align}
and a spectator component
\begin{align}
\mathcal{B}_s(3e^1)
&=
\begin{Bmatrix}
\ket{1,\downarrow,0_\lambda} & \overset{\mathrm{VC}}{\Leftrightarrow} & \ket{-1,\downarrow,0_\lambda}
\vspace{0.2cm}
\\
\Updownarrow \mathrm{cZee} & & \Updownarrow \mathrm{cZee}
\vspace{0.2cm}
\\
\ket{1,\uparrow,1_\lambda} & \overset{\mathrm{VC}}{\Leftrightarrow} & \ket{-1,\uparrow,1_\lambda}
\end{Bmatrix}
\label{eq.spectator_basis_e1}
\vspace{0.2cm}
\\
\mathcal{B}_s(2e^3)
&=
\small{
\begin{Bmatrix}
\ket{1_\downarrow,-1_\uparrow,-1_\downarrow,0_z} & \overset{\mathrm{VC}}{\Leftrightarrow} & \ket{1_\uparrow,1_\downarrow,-1_\downarrow,0_z}
\vspace{0.2cm}
\\
\Updownarrow \mathrm{cZee} & & \Updownarrow \mathrm{cZee}
\vspace{0.2cm}
\\
\ket{1_\uparrow,-1_\uparrow,-1_\downarrow,1_z} & \overset{\mathrm{VC}}{\Leftrightarrow} & \ket{1_\uparrow,1_\downarrow,-1_\uparrow,1_z}
\end{Bmatrix}
}\normalsize
\label{eq.spectator_basis_e3}
\end{align} 
in line with terminology introduce in Ref.\cite{fischer2025zeeman}. In both cases vibronic coupling (VC) and cavity Zeeman interaction (cZee) is explicitly highlighted and we stress that there is no interaction between polariton and spectator subspaces. Accordingly, the cavity-modified RJT Hamiltonian in Eq.\eqref{eq.cavity_rjt_hamilton} takes a block-diagonal matrix representation 
\begin{align}
\underline{\underline{H}}_\mathrm{cRJT}
&=
\begin{pmatrix}
\underline{\underline{H}}_p & \underline{\underline{0}} \\
\underline{\underline{0}}   & \underline{\underline{H}}_s
\end{pmatrix}
\quad,
\label{eq.crjt_hamilton_basis}
\end{align}
with the polariton Hamiltonian being explicitly given by
\begin{widetext}
\begin{align}
\underline{\underline{H}}_p
=
\begin{pmatrix}
-
\dfrac{\xi}{2}
\mp
(1\mp\frac{g_e}{2})
\mu_B
B_z
& 
F\rho 
&
0
&
\dfrac{g_0\,g_e\mu_B}{2c}
\sqrt{\dfrac{\hbar\omega_c}{2}}
\vspace{0.2cm}
\\
F\rho 
& 
\dfrac{\xi}{2}
\pm
(1\pm\frac{g_e}{2})
\mu_B 
B_z
&
\dfrac{g_0\,g_e\mu_B}{2c}
\sqrt{\dfrac{\hbar\omega_c}{2}}
&
0
\vspace{0.2cm}
\\
0
&
\dfrac{g_0\,g_e\mu_B}{2c}
\sqrt{\dfrac{\hbar\omega_c}{2}}
&
-
\dfrac{\xi}{2} 
+
\hbar\omega_c
\pm
(1\mp\frac{g_e}{2})
\mu_B
B_z
&
F\rho
\vspace{0.2cm}
\\
\dfrac{g_0\,g_e\mu_B}{2c}
\sqrt{\dfrac{\hbar\omega_c}{2}}
&
0
&
F\rho 
&
\dfrac{\xi}{2} 
+
\hbar\omega_c
\mp
(1\pm\frac{g_e}{2})
\mu_B 
B_z
\end{pmatrix}
\,.
\label{eq.polariton_hamilton}
\end{align}
\end{widetext}
where underlying basis states are ordered clockwise in Eqs.\eqref{eq.polariton_basis_e1} and \eqref{eq.polariton_basis_e3}, while sign convention on the diagonal is as before, \textit{i.e.}, upper signs reflect the single-particle and lower signs the single-hole basis. In Figs.\ref{fig.crjt_polariton_spectator_dispersion}a and c, we depict energies of zero-order basis states as well as their dispersion with respect to the external classical B-field amplitude.   
We realize now that the lower $2\times2$-block subject to vibronic coupling is equivalent to $\underline{\underline{H}}_\uparrow$ in Eq.\eqref{eq.rjt_hamilton_basis_up}. Vibronic coupling is maximized at a classical B-field amplitude 
\begin{align}
B^F_z
&=
\dfrac{\xi}{2\mu_B}
\quad,
\end{align}
which holds also for single-hole polariton and single-particle spectator problems illustrated in Figs.\ref{fig.crjt_polariton_spectator_dispersion}c and b. The central $2\times2$-block of Eq.\eqref{eq.polariton_hamilton} leads to cavity-Zeeman-mediated spin-polariton formation maximized at a classical B-field amplitude (\textit{cf.} Figs.\ref{fig.crjt_polariton_spectator_dispersion}a and c) 
\begin{align}
B^\star_z
=
\dfrac{
\hbar\omega_c
-
\xi}{g_e
\mu_B }
\quad.
\end{align}
The spectator problem is determined by the lower block of Eq.\eqref{eq.crjt_hamilton_basis}, respectively denoted as spectator Hamiltonian, which reads explicitly
\begin{widetext}
\begin{align}
\underline{\underline{H}}_s
=
\begin{pmatrix}
-
\dfrac{\xi}{2}
\pm
(1\mp\frac{g_e}{2})
\mu_B
B_z
& 
F\rho 
&
0
&
-
\dfrac{g_0\,g_e\mu_B}{2c}
\sqrt{\dfrac{\hbar\omega_c}{2}}
\vspace{0.2cm}
\\
F\rho
& 
\dfrac{\xi}{2}
\mp
(1\pm\frac{g_e}{2})
\mu_B 
B_z
&
-
\dfrac{g_0\,g_e\mu_B}{2c}
\sqrt{\dfrac{\hbar\omega_c}{2}}
&
0
\vspace{0.2cm}
\\
0
&
-
\dfrac{g_0\,g_e\mu_B}{2c}
\sqrt{\dfrac{\hbar\omega_c}{2}}
&
-
\dfrac{\xi}{2} 
+
\hbar\omega_c
\mp
(1\mp\frac{g_e}{2})
\mu_B
B_z
&
F\rho 
\vspace{0.2cm}
\\
-
\dfrac{g_0\,g_e\mu_B}{2c}
\sqrt{\dfrac{\hbar\omega_c}{2}}
&
0
&
F\rho 
&
\dfrac{\xi}{2} 
+
\hbar\omega_c
\pm
(1\pm\frac{g_e}{2})
\mu_B 
B_z
\end{pmatrix}
\,,
\label{eq.spectator_hamilton}
\end{align}
\end{widetext}
where related basis states are given in Eqs.\eqref{eq.spectator_basis_e1} and \eqref{eq.spectator_basis_e3}. We find the lower $2\times 2$-block to be equivalent to $\underline{\underline{H}}_\downarrow$ in Eq.\eqref{eq.rjt_hamilton_basis_down} and there exists no classical B-field amplitude leading to spin-polariton formation via the central $2\times 2$-block, which motivated the notion of a \textit{spectator} basis.\cite{fischer2025zeeman}.

We have now established a model Hamiltonian for a cavity-modified relativistic $E\times e$-Jahn-Teller problem in presence of classical and quantized Zeeman interactions, which generalizes molecular $\uparrow$- and $\downarrow$-sectors (\textit{cf.} Eqs.\eqref{eq.rjt_hamilton_basis_up} and \eqref{eq.rjt_hamilton_basis_down}) to polariton and spectator problems. The respective ``ground states'' accordingly reflect the cavity-modified Kramers pair in absence of an external classical B-field and accordingly determine the Zeeman splitting as well as the electronic g-factor. In the remainder of this work, we shall analyse cavity-Zeeman-induced signatures by means of effective Hamiltonian theory. 

\section{Cavity-Modified RJT Effect}
\label{sec.cavity_modified_rjt}
We are ultimately interested in understanding cavity-induced corrections of the ``molecular'' electronic g-factor under weak and strong SOC (\textit{cf.} Eqs.\eqref{eq.geff_weakSOC} and \eqref{eq.geff_strongSOC}) as defined in the weak classical B-field limit ($0<B_z\ll B^\star_z$). This assumption will allows us here to simplify polariton and spectator subspaces by neglecting cavity-Zeeman interactions between energetically well-separated ground and highest-lying excited states to obtain effective three dimensional problems. This approximation is illustrated via a green box (\xmark) in Fig.\ref{fig.crjt_polariton_spectator_dispersion}, where we assume 
\begin{align}
\hbar\omega_c
>
\mu_B B_z,
\xi,
F\rho
\quad.
\label{eq.model_energy_scales}
\end{align}
We will then solve the vibronic coupling problem in the energetically lower-lying two-dimensional subspace \textit{exactly} providing access to eigenvalues of the molecular problem discussed in Sec.\ref{sec.rjt_hamilton}. Related eigenstates are now subject to ``vibronic coupling-dressed'' cavity-Zeeman interactions with the remaining excited state (\textit{cf.} Fig.\ref{fig.crjt_polariton_spectator_dispersion}, purple box (``Exact'')). Motivated by the observation that this excited state is energetically well-separated from the ground state for weak B-fields (\textit{cf.} Fig.\ref{fig.crjt_polariton_spectator_dispersion}, blue box (``QDPT'')), we consequently exploit quasi-degenerate perturbation theory (QDPT) to obtain cavity-corrected ground states for polariton and spectator problems.

\subsection{Effective Hamiltonian in Weak B-Field Limit}
\label{sec.weak_B_field_limit}
In the weak B-field limit, we neglect energetically highest-lying states in $\mathcal{B}_p(X)$ and $\mathcal{B}_s(X)$ to obtain three dimensional effective polariton and spectator Hamiltonians (\textit{cf.} Appendix \ref{sec.details_eff_hamilton})
\begin{align}
\underline{\underline{H}}_i
\overset{0<B_z\ll B^\star_z}{\Longrightarrow}
\underline{\underline{\tilde{H}}}_i
\quad,
\quad
i
=
p,s
\quad,
\label{eq.polariton_hamilton_eff}
\end{align}
Exact diagonalization of the lower-lying $2\times2$-subblocks recovers eigenvalues of $\underline{\underline{H}}_\uparrow$ and $\underline{\underline{H}}_\downarrow$, respectively, and is accomplished via a unitary transformation acting on the truncated model space
\begin{align}
\underline{\underline{\tilde{U}}}_i(X)
&=
\begin{pmatrix}
\cos\frac{\varphi_i(X)}{2} & -\sin\frac{\varphi_i(X)}{2} & 0
\vspace{0.2cm}
\\
\sin\frac{\varphi_i(X)}{2} & \cos\frac{\varphi_i(X)}{2}  & 0
\vspace{0.2cm}
\\
0 & 0 & 1
\end{pmatrix}
\,,\,
i
=
p,s
\quad,
\label{eq.unitary_rotation}
\end{align}
with configuration-dependent transformation angles
\begin{align}
\begin{matrix}
\varphi_p(3e^1)
&=
-
\arctan\left(\dfrac{2F\rho}{2\mu_B B_z+\xi}\right)
=
\varphi_s(2e^3)
\vspace{0.2cm}
\\
\varphi_s(3e^1)
&=
\arctan\left(\dfrac{2F\rho}{2\mu_B B_z-\xi}\right)
=
\varphi_p(2e^3)
\end{matrix}
\quad.
\label{eq.rotation_angles}
\end{align}
The approximate polariton and spectator Hamiltonians are then given by
\begin{align}
(
\underline{\underline{\tilde{U}}}^\dagger_i
\underline{\underline{H}}_i
\underline{\underline{\tilde{U}}}_i
)
(X)
=
\begin{pmatrix}
E^i_0(X)
& 
0 
&
v^i_{02}(X)
\vspace{0.2cm}
\\
0 
& 
E^i_1(X)
&
v^i_{12}(X)
\vspace{0.2cm}
\\
v^i_{02}(X)
&
v^i_{12}(X)
&
E^i_2(X)
\end{pmatrix}
\,,
\label{eq.transform_3x3_hamilton_l1}
\end{align}
where $E^i_0(X),E^i_1(X)$ and $E^i_2(X)$ are explicitly given in Appendices \ref{sec.details_rjt} and \ref{sec.details_eff_hamilton}. Matrix elements for the ``vibronic coupling''-dressed cavity-Zeeman interactions read
\begin{align}
v^i_{02}(X)
&=
\dfrac{g_0\,g_e\mu_B}{2c}
\sqrt{\dfrac{\hbar\omega_c}{2}}
\sin\frac{\varphi_i(X)}{2}
\quad,
\label{eq.dressed_cavity_zeeman_m}
\vspace{0.2cm}
\\
v^i_{12}(X)
&=
\dfrac{g_0\,g_e\mu_B}{2c}
\sqrt{\dfrac{\hbar\omega_c}{2}}
\cos\frac{\varphi_i(X)}{2}
\quad,
\label{eq.dressed_cavity_zeeman_p}
\end{align}
where we shall restrict our attention in the following to the interaction, $v^i_{02}(X)$, involving the ground state. Weak and strong SOC approximations of the interaction matrix element for the single-particle scenario read
\small{
\begin{align}
v^i_{02}(3e^1)
&=
\frac{g_0\,g_e\mu_B}{2c}
\sqrt{\frac{\hbar\omega_c}{2}}
(
-
\dfrac{1}{\sqrt{2}}
\pm
\dfrac{\mu_B B_z}{2\sqrt{2}F\rho}
)
+
\mathcal{O}(B^2_z,\xi^2)
,
\label{vc_dressed_czee_weakSOC}
\vspace{0.2cm}
\\
v^i_{02}(3e^1)
&=
\frac{g_0\,g_e\mu_B}{2c}
\sqrt{\frac{\hbar\omega_c}{2}}
(
-
\dfrac{F\rho}{\xi}
\pm
\dfrac{2F\rho}{\xi}
\mu_B B_z
)
+
\mathcal{O}(B^2_z,F^2)
,
\label{vc_dressed_czee_strongSOC}
\end{align}
}\normalsize
for polariton (upper sign) and spectator (lower sign) scenario problems. The single-hole expressions follow by symmetry from Eq.\eqref{eq.rotation_angles}. Notably, in line with Ref.\cite{fischer2025zeeman}, we find also here an interplay of classical and cavity magnetic field components. 

\subsection{Kramers Pair via QDPT}
\label{sec.kramers_qdpt}
We obtain approximate ground state energies of polariton and spectator problems at second-order QDPT as
\begin{align}
\tilde{E}^i_0(X)
&=
E^i_0(X)
-
\dfrac{v^i_{02}(X)v^i_{02}(X)}{\Delta_{02}}
\quad,
\label{eq.qdpt_states}
\end{align}
with $\Delta_{02}\approx\hbar\omega_c$ in the weak B-field limit (\textit{cf.} Appendix \ref{sec.details_eff_hamilton}). Note, the second-order correction obtained from Eqs.\eqref{vc_dressed_czee_weakSOC} contains a term quadratic in the B-field amplitude, which we neglect due to the weak B-field approximation. 

For the polariton problem, we consequently find approximate ground state energies 
\begin{multline}
\tilde{E}^p_0
=
\dfrac{1}{2}
\left(
g_e
\mp
\dfrac{\xi}{F\rho}
\pm
\dfrac{g^2_0 g^2_e\mu^2_B}{16c^2}
\dfrac{1}{F\rho}
\right)
\mu_B
B_z
\\
-
F\rho
-
\dfrac{\xi^2}{8F\rho}
-
\dfrac{g^2_0 g^2_e\mu^2_B}{16c^2}
+
\mathcal{O}(B^2_z,\xi^3)
\quad,
\end{multline}
for the weak SOC regime, while the strong SOC counterparts read
\begin{multline}
\tilde{E}^p_0
=
\dfrac{1}{2}
\left(
g_e
\mp
2
\pm
\dfrac{4F^2\rho^2}{\xi^2}
\pm
\dfrac{g^2_0 g^2_e\mu^2_B}{2c^2}
\dfrac{F^2\rho^2}{\xi^3}
\right)
\mu_B
B_z
\\
-
\dfrac{\xi}{2}
-
\dfrac{F^2\rho^2}{\xi}
-
\dfrac{g^2_0 g^2_e\mu^2_B}{8c^2}
\dfrac{F^2\rho^2}{\xi^2}
+
\mathcal{O}(B^2_z,F^3)
\quad.
\end{multline}
Turning to the spectator problem, energies in the weak SOC regime take the form
\begin{multline}
\tilde{E}^s_0
=
-
\dfrac{1}{2}
\left(
g_e
\mp
\dfrac{\xi}{F\rho}
\pm
\dfrac{g^2_0 g^2_e\mu^2_B}{16c^2}
\dfrac{1}{F\rho}
\right)
\mu_B
B_z
\\
-
F\rho
-
\dfrac{\xi^2}{8F\rho}
-
\dfrac{g^2_0 g^2_e\mu^2_B}{16c^2}
+
\mathcal{O}(B^2_z,\xi^3)
\quad,
\end{multline}
and for strong SOC, one obtains
\begin{multline}
\tilde{E}^s_0
=
-
\dfrac{1}{2}
\left(
g_e
\mp
2
\pm
\dfrac{4F^2\rho^2}{\xi^2}
\pm
\dfrac{g^2_0 g^2_e\mu^2_B}{2c^2}
\dfrac{F^2\rho^2}{\xi^3}
\right)
\mu_B
B_z
\\
-
\dfrac{\xi}{2}
-
\dfrac{F^2\rho^2}{\xi}
-
\dfrac{g^2_0 g^2_e\mu^2_B}{8c^2}
\dfrac{F^2\rho^2}{\xi^2}
+
\mathcal{O}(B^2_z,F^3)
\quad.
\end{multline}
In both SOC regimes, we find the doubly degenerate Kramers pair with energies $\tilde{E}^p_0=\tilde{E}^s_0$ for a vanishing external B-field amplitude. We obtain cavity-induced corrections independent of the cavity frequency due to the weak B-field approximation where the prefactor proportional to $g^2_0$ has a unit of energy (\textit{cf.} Appendix \ref{sec.dim_anal}). Moreover, for weak SOC we find a correction exclusively determined by the cavity-Zeeman interaction whereas the strong SOC modification reflects a complex interplay of vibronic, spin-orbit and cavity Zeeman interactions.

\subsection{Cavity-Modified Electronic g-factor}
\label{sec.cavity_zeeman_rjt}
We are finally in the position to discuss cavity-modified effective electronic g-factors for single-particle and single-hole scenarios, which straightforwardly follow from Sec.\ref{sec.kramers_qdpt} as
\begin{align}
\tilde{g}_\mathrm{eff}
&=
g_e
\mp
\dfrac{\xi}{F\rho}
\pm
\dfrac{g^2_0 g^2_e\mu^2_B}{8c^2}
\dfrac{1}{F\rho}
+
\mathcal{O}(\xi^3)
\quad,
\label{eq.geff_weakSOC_cavity}
\vspace{0.2cm}
\\
\tilde{g}_\mathrm{eff}
&=
g_e
\mp
2
\pm
4
\dfrac{F^2\rho^2}{\xi^2}
\pm
\dfrac{g^2_0 g^2_e\mu^2_B}{c^2}
\dfrac{F^2\rho^2}{\xi^3}
+
\mathcal{O}(F^3)
\,.
\label{eq.geff_strongSOC_cavity}
\end{align}
We immediately recover molecular results previously presented in Eqs.\eqref{eq.geff_weakSOC} and \eqref{eq.geff_strongSOC} besides two distinct cavity-induced corrections to be discussed in the following.
\begin{figure}[hbt!]
\includegraphics[scale=1.0]{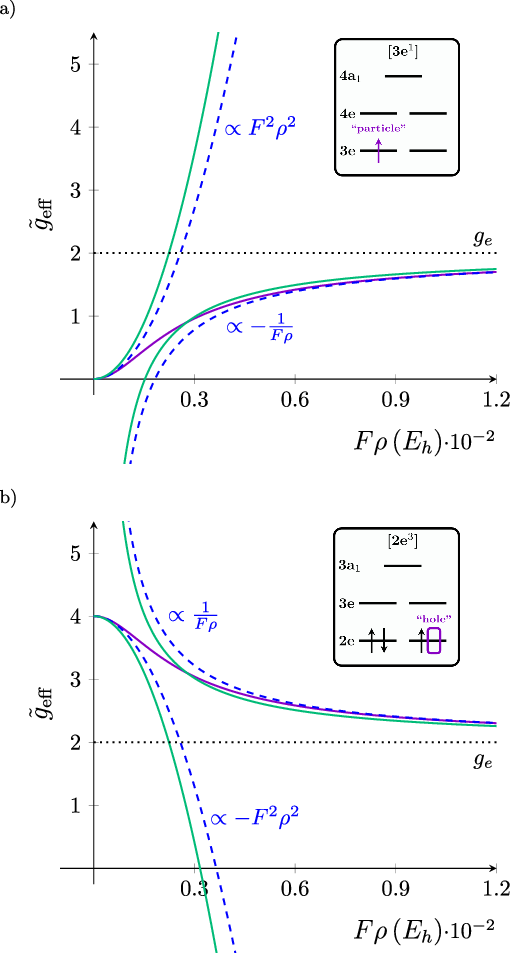}
\caption{Effective electronic g-factors for (a) single-particle and (b) single-hole scenarios as function of vibronic coupling strength, $F\rho$, with SOC strength, $\xi=800\,\mathrm{cm}^{-1}=3.65\times10^{-3}E_h$ and effective light-matter interaction constant, $\tilde{g}_0=\sqrt{N}g_0$, for a collection of $N=10^5$ molecules and $g_0=0.03\sqrt{E_h}/ea_0$\cite{fischer2025cbocc}. Dashed blue and bold green graphs reflect molecular and cavity-corrected approximations, respectively, for weak SOC $(\xi\ll F\rho)$ and strong SOC $(\xi\gg F\rho)$ regimes.}
\label{fig.crjt_cavity_geff_dispersion}
\end{figure}

In Fig.\ref{fig.crjt_cavity_geff_dispersion}, we depict effective electronic g-factors of single-particle $[3e^1]$ and single-hole $[2e^3]$ systems in the weak B-field limit as function of vibronic coupling strength, $F\rho$. We chose $\xi=800\,\mathrm{cm}^{-1}=3.65\times10^{-3}E_h$ for the SOC strength in analogy to single-particle and single-hole Mo-complexes studied in Refs.\cite{sharma2017,mcnaughton2010}. The molecular result in Eq.\eqref{eq.rjt_gfac} converges to antiferromagnetically quenched $(g_e-2)$ and ferromagnetically enhanced $(g_e+2)$ limits as $F\to0$ (strong SOC, $\xi\gg F\rho$). The ``spin-only'' regime determined by the free electronic g-factor, $g_e$, is reached for both scenarios as $F\to\infty$ (weak SOC, $\xi\ll F\rho$) since orbital angular momentum is then quenched by vibronic coupling. 

We turn now to cavity-induced modifications in weak and strong SOC regimes for a collective light-matter interaction scenario with effective coupling strength, $\tilde{g}_0=\sqrt{N}g_0$, for $N=10^5$ molecules and $g_0=0.03\sqrt{E_h}/ea_0$. Both single-particle and single-hole scenarios are subject to a more pronounced cavity-Zeeman modification in the weak SOC regime, which is rationalized by comparing higher-order corrections in Eqs.\eqref{eq.geff_weakSOC_cavity} and \eqref{eq.geff_strongSOC_cavity} equally written as
\begin{align}
\Delta\tilde{g}_\mathrm{eff}
&=
\mp
\left(
\xi
-
\dfrac{g^2_0 g^2_e\mu^2_B}{8c^2}
\right)
\dfrac{1}{F\rho}
+
\mathcal{O}(\xi^3)
\quad,
\label{eq.dgeff_weakSOC_cavity}
\vspace{0.2cm}
\\
\Delta\tilde{g}_\mathrm{eff}
&=
\pm
\left(
4
+
\dfrac{g^2_0 g^2_e\mu^2_B}{c^2}
\dfrac{1}{\xi}
\right)
\dfrac{F^2\rho^2}{\xi^2}
+
\mathcal{O}(F^3)
\quad.
\label{eq.dgeff_strongSOC_cavity}
\end{align}
The weak SOC regime translates into a small SOC strength, $\xi$, such that the cavity-Zeeman correction in Eq.\eqref{eq.dgeff_weakSOC_cavity} can become relevant relative to the leading-order ``molecular'' correction linear in $\xi$. In the strong SOC regime in contrast, the cavity-Zeeman interaction is quenched as $\xi^{-1}$ (\textit{cf.} Eq.\eqref{eq.dgeff_strongSOC_cavity}) for large SOC strengths $\xi$. Our analysis thus suggests that spin systems subject to weak SOC might be more prone to cavity-induced corrections due to a more prominent Zeeman interaction between electronic spins and the cavity magnetic field component effectively quenched in the strong SOC regime. We can finally distinguish cavity-induced corrections for single-particle and single-hole scenarios in the weak SOC regime qualitatively as $\tilde{g}_\mathrm{eff}$ increases for the former and decreases for the latter.

\section{Conclusion}
\label{sec.conclusion}
We studied the cavity-Zeeman effect in a relativistic $E\times e$-Jahn-Teller model of single-particle/-hole systems realized by trigonal symmetric transition metal complexes with a molybdenum (Mo) center\cite{sharma2017,mcnaughton2010} under strong light-matter coupling. To this end, we introduced a generalized effective RJT model Hamiltonian accounting for cavity Zeeman interactions of electronic spin and cavity magnetic field components beyond the commonly employed dipole approximation extending our previous related work\cite{fischer2025zeeman} to a realistically complex molecular model environment. We derived analytic expressions for cavity-modified electronic g-factors subject to spin-orbit coupling (SOC), vibronic coupling and the cavity-Zeeman interaction by means of effective Hamiltonian theory in the weak classical B-field limit. 

In the absence of a classical magnetic field, we obtain a cavity-modified Kramers pair subject to leading-order corrections determined by vibronic, spin-orbit and cavity Zeeman interactions with identical results for single-particle and single-hole scenarios. The effective electronic g-factor is found to be prone to cavity-induced corrections in the weak SOC regime, where single-particle and single-hole systems acquire additional contributions with alternating sign. In contrast, the cavity-Zeeman interaction is effectively quenched in the strong SOC regime for both scenarios. Our study highlights the rich properties of molecular spin systems under strong light-matter coupling potentially approachable by means of electron paramagnetic resonance (EPR) spectropscopy as discussed previously\cite{fischer2025zeeman}.

A natural extension of the present molecular model could account for electronic excited states inducing interactions with additional components of orbital and spin angular momentum. Another aspect relates to the deviation from the ``perfect'' atomic orbital picture assumed here due to covalent ligand-metal interactions. From the light-matter perspective, higher-order spin-cavity interactions beyond the cavity Zeeman effect can become relevant, which are significantly less explored than related electronic and vibrational strong coupling problems. This might provide additional insights into how strong light-matter coupling could be exploited to shape spin properties of molecular systems as transition metal complexes or radicals and their specific chemistry.

\section*{Conflict of Interest}
The authors have no conflicts to disclose.

\section*{Data Availability Statement}
Data sharing is not applicable to this article as no new data were created or analysed in this study.

\section*{Acknowledgements}
E.W. Fischer acknowledges funding by the Deutsche Forschungsgemeinschaft (DFG, German Research Foundation) through DFG project 536826332.

\renewcommand{\thesection}{}
\section*{Appendix}

\setcounter{equation}{0}
\renewcommand{\theequation}{\thesubsection.\arabic{equation}}
\subsection{Details of the RJT Hamiltonian}
\label{sec.details_rjt_hamilton}

\subsubsection{Single-Particle Scenario}
The single-particle SOC operator reads
\begin{align}
\hat{H}_\mathrm{soc}
&=
\xi
\hat{L}_z
\hat{S}_z
\quad,
\label{eq.soc_hamilton_e1}
\end{align}
with SOC coupling strength, $\xi$, besides single-electron orbital and spin angular momentum operators
\begin{align}
\hat{L}_z
&=
\hbar
(
\ket{1}
\bra{1}
-
\ket{-1}
\bra{-1}
)
\quad,
\label{eq.ang_op}
\vspace{0.2cm}
\\
\hat{S}_z
&=
\dfrac{\hbar}{2}
\left(
\ket{\uparrow}
\bra{\uparrow}
-
\ket{\downarrow}
\bra{\downarrow}
\right)
\quad,
\label{eq.spin_op}
\end{align}
acting on orbital and spin angular momentum eigenstates in $z$-axis projection, $\ket{\pm 1}$ and $\ket{\uparrow},\ket{\downarrow}$, respectively. We omit $x$- and $y$-components of the orbital angular momentum operator, which vanish for relevant $d_{xz},d_{yz}$ orbitals. The classical Zeeman interaction reads
\begin{align}
\hat{H}_\mathrm{Zee}
&=
\dfrac{\mu_B}{\hbar}
\left(
\hat{L}_z
+
g_e
\hat{S}_z
\right)
B_z
\quad,
\label{eq.zeeman_op}
\end{align}
with Bohr Magneton, $\mu_B$, electronic g-factor for the free electron, $g_e$, and classical B-field amplitude, $B_z$, for a field aligned parallel to the $z$-axis. Finally, the linear vibronic interaction reads for the single-particle scenario
\begin{align}
\hat{H}_\mathrm{vc}
&=
F
\left(
\rho 
e^{+\mathrm{i}\phi}
\ket{1}
\bra{-1}
+
\rho 
e^{-\mathrm{i}\phi}
\ket{-1}
\bra{1}
\right)
\quad,
\label{eq.vc_op}
\end{align}
with vibronic coupling strength, $F$, and complex valued normal-mode displacements, $Q_\pm=\rho e^{\pm\mathrm{i}\phi}$, determined by polar coordinates, $\rho$ and $\phi$, respectively. Matrix elements of SOC and Zeeman operators with respect to the basis in Eq.\eqref{eq.rjt_basis_e1} are given by
\begin{align}
\begin{matrix}
\braket{
1,\uparrow
\vert
\hat{H}_\mathrm{soc}
\vert
1,\uparrow}
=
\dfrac{\xi}{2}
\quad,
\vspace{0.2cm}
\\
\braket{
1,\downarrow
\vert
\hat{H}_\mathrm{soc}
\vert
1,\downarrow}
=
-\dfrac{\xi}{2}
\quad,
\vspace{0.2cm}
\\
\braket{
-1,\uparrow
\vert
\hat{H}_\mathrm{soc}
\vert
-1,\uparrow}
=
-\dfrac{\xi}{2}
\quad,
\vspace{0.2cm}
\\
\braket{
-1,\downarrow
\vert
\hat{H}_\mathrm{soc}
\vert
-1,\downarrow}
=
\dfrac{\xi}{2}
\quad,
\end{matrix}
\label{eq.soc_matrix_d1}
\end{align}
and 
\begin{align}
\begin{matrix}
\braket{
1,\uparrow
\vert
\hat{H}_\mathrm{Zee}
\vert
1,\uparrow}
=
(
1
+
\frac{g_e}{2}
)
\mu_B
B_z
\quad,
\vspace{0.2cm}
\\
\braket{
1,\downarrow
\vert
\hat{H}_\mathrm{Zee}
\vert
1,\downarrow}
=
(
1
-
\frac{g_e}{2}
)
\mu_B
B_z
\quad,
\vspace{0.2cm}
\\
\braket{
-1,\uparrow
\vert
\hat{H}_\mathrm{Zee}
\vert
-1,\uparrow}
=
-
(
1
-
\frac{g_e}{2}
)
\mu_B
B_z
\quad,
\vspace{0.2cm}
\\
\braket{
-1,\downarrow
\vert
\hat{H}_\mathrm{Zee}
\vert
-1,\downarrow}
=
-
(
1
+
\frac{g_e}{2}
)
\mu_B
B_z
\quad,
\end{matrix}
\label{eq.zee_matrix_d1}
\end{align}
while all off-diagonal elements vanish identically.

\subsubsection{Single-Hole Scenario}
For the single-hole scenario, we consider three electrons in a doubly-degenerate d-orbital ($e$) with SOC and Zeeman operators 
\begin{align}
\hat{H}_\mathrm{soc}
&=
\xi
\sum^{N_e}_i
\hat{L}^i_z
\hat{S}^i_z
\quad,
\vspace{0.2cm}
\\
\hat{H}_\mathrm{Zee}
&=
\dfrac{\mu_B}{\hbar}
\sum^{N_e}_i
\left(
\hat{L}^i_z
+
g_e
\hat{S}^i_z
\right)
B_z
\quad,
\end{align}
where $N_e=3$ while $\hat{L}^i_z$ and $\hat{S}^i_z$ are similar to Eqs.\eqref{eq.ang_op} and \eqref{eq.spin_op}, respectively. The vibronic coupling interaction reads for the $\uparrow$-sector
\begin{align}
\hat{H}_\mathrm{vc}
&=
F
\left(
\rho 
e^{+\mathrm{i}\phi}
\ket{1_\uparrow,1_\downarrow,-1_\uparrow}
\bra{1_\uparrow,-1_\uparrow,-1_\downarrow}
+
c.c.
\right)
\,,
\end{align}
and
\begin{align}
\hat{H}_\mathrm{vc}
&=
F
\left(
\rho 
e^{+\mathrm{i}\phi}
\ket{1_\uparrow,1_\downarrow,-1_\downarrow}
\bra{1_\downarrow,-1_\uparrow,-1_\downarrow}
+
c.c.
\right)
\,,
\end{align}
for the $\downarrow$-sector. Matrix elements read
\begin{align}
\begin{matrix}
\braket{
1_\uparrow,1_\downarrow,-1_\uparrow
\vert
\hat{H}_\mathrm{soc}
\vert
1_\uparrow,1_\downarrow,-1_\uparrow
}
=
-
\dfrac{\xi}{2}
\vspace{0.2cm}
\\
\braket{
1_\uparrow,-1_\uparrow,-1_\downarrow
\vert
\hat{H}_\mathrm{soc}
\vert
1_\uparrow,-1_\uparrow,-1_\downarrow
}
=
\dfrac{\xi}{2}
\vspace{0.2cm}
\\
\braket{
1_\uparrow,1_\downarrow,-1_\downarrow
\vert
\hat{H}_\mathrm{soc}
\vert
1_\uparrow,1_\downarrow,-1_\downarrow
}
=
\dfrac{\xi}{2}
\vspace{0.2cm}
\\
\braket{
1_\downarrow,-1_\uparrow,-1_\downarrow
\vert
\hat{H}_\mathrm{soc}
\vert
1_\downarrow,-1_\uparrow,-1_\downarrow
}
=
-
\dfrac{\xi}{2}
\end{matrix}
\quad,
\end{align}
and 
\begin{align}
\begin{matrix}
\braket{
1_\uparrow,1_\downarrow,-1_\uparrow
\vert
\hat{H}_\mathrm{Zee}
\vert
1_\uparrow,1_\downarrow,-1_\uparrow
}
=
(
1
+
\frac{g_e}{2}
)
\mu_B
B_z
\vspace{0.2cm}
\\
\braket{
1_\uparrow,-1_\uparrow,-1_\downarrow
\vert
\hat{H}_\mathrm{Zee}
\vert
1_\uparrow,-1_\uparrow,-1_\downarrow
}
=
-
(
1
-
\frac{g_e}{2}
)
\mu_B
B_z
\vspace{0.2cm}
\\
\braket{
1_\uparrow,1_\downarrow,-1_\downarrow
\vert
\hat{H}_\mathrm{Zee}
\vert
1_\uparrow,1_\downarrow,-1_\downarrow
}
=
(
1
-
\frac{g_e}{2}
)
\mu_B
B_z
\vspace{0.2cm}
\\
\braket{
1_\downarrow,-1_\uparrow,-1_\downarrow
\vert
\hat{H}_\mathrm{Zee}
\vert
1_\downarrow,-1_\uparrow,-1_\downarrow
}
=
-
(
1
+
\frac{g_e}{2}
)
\mu_B
B_z
\end{matrix}
.
\end{align}

\setcounter{equation}{0}
\renewcommand{\theequation}{\thesubsection.\arabic{equation}}
\subsection{Details of the RJT Model}
\label{sec.details_rjt}
\subsubsection{Ground State}
The energetically lower-lying states of $\underline{\underline{H}}_\uparrow$ and $\underline{\underline{H}}_\downarrow$ (\textit{cf.} Eqs.\eqref{eq.rjt_hamilton_basis_up} and \eqref{eq.rjt_hamilton_basis_down}) correspond to $E^i_0(X)$ in Eq.\eqref{eq.transform_3x3_hamilton_l1} given by
\begin{align}
E^i_0(X)
&=
E^p_0(X)
=
E^\uparrow_0
\quad,
\vspace{0.2cm}
\\
E^i_0(X)
&=
E^s_0(X)
=
E^\downarrow_0
\quad.
\end{align}
Explicitly, we have
\begin{align}
E^\uparrow_0
&=
\dfrac{g_e
\mu_B}{2}
B_z
-
\sqrt{
\left(
\mu_B
B_z
\pm
\frac{\xi}{2}
\right)^2
+
F^2
\rho^2
}
\,,
\end{align}
and
\begin{align}
E^\downarrow_0
&=
-
\dfrac{g_e
\mu_B}{2}
B_z
-
\sqrt{
\left(
\mu_B
B_z
\mp
\frac{\xi}{2}
\right)^2
+
F^2
\rho^2
}
\,,
\end{align}
where upper signs reflect single-particle $([3e^1])$ and lower signs single-hole $([2e^3])$ scenarios. The weak $B_z$-field limit is determined by linear Taylor expansions
\begin{multline}
E^\uparrow_0
=
-
\dfrac{1}{2}
\sqrt{
\xi^2
+
4
F^2
\rho^2
}
\\
+
\left(
\dfrac{g_e\mu_B}{2}
\mp
\dfrac{\mu_B\xi}{\sqrt{\xi^2+4F^2\rho^2}}
\right)
B_z
+
\mathcal{O}(B^2_z)
\quad,
\end{multline}
and
\begin{multline}
E^\downarrow_0
=
-
\dfrac{1}{2}
\sqrt{
\xi^2
+
4
F^2
\rho^2
}
\\
+
\left(
-
\dfrac{g_e\mu_B}{2}
\pm
\dfrac{\mu_B\xi}{\sqrt{\xi^2+4 F^2\rho^2}}
\right)
B_z
+
\mathcal{O}(B^2_z)
\quad,
\end{multline}
from which we obtain $\Delta_\mathrm{Zee}$ in Eq.\eqref{eq.rjt_zeeman_split} with $g_\mathrm{eff}$ given in Eq.\eqref{eq.rjt_gfac}, respectively. We expand now both expressions around $\xi=0$ and $F=0$ to obtain approximations for weak $(\xi\ll F\rho)$ and strong $(\xi\gg F\rho)$ SOC regimes. For the $\uparrow$-sector, we find
\begin{multline}
E^\uparrow_0
=
\dfrac{1}{2}
\left(
g_e
\mp
\dfrac{\xi}{F\rho}
\right)
\mu_B
B_z
\\
-
F\rho
-
\dfrac{\xi^2}{8F\rho}
+
\mathcal{O}(B^2_z,\xi^3)
\quad,
\end{multline}
and
\begin{multline}
E^\uparrow_0
=
\dfrac{1}{2}
\left(
g_e
\mp
2
\pm
\dfrac{4F^2\rho^2}{\xi^2}
\right)
\mu_B
B_z
\\
-
\dfrac{\xi}{2}
-
\dfrac{F^2\rho^2}{\xi}
+
\mathcal{O}(B^2_z,F^3)
\quad,
\end{multline}
while the $\downarrow$-sector is determined by
\begin{multline}
E^\downarrow_0
=
-
\dfrac{1}{2}
\left(
g_e
\mp
\dfrac{\xi}{F\rho}
\right)
\mu_B
B_z
\\
-
F\rho
-
\dfrac{\xi^2}{8F\rho}
+
\mathcal{O}(B^2_z,\xi^3)
\quad,
\end{multline}
and 
\begin{multline}
E^\downarrow_0
=
-
\dfrac{1}{2}
\left(
g_e
\mp
2
\pm
\dfrac{4F^2\rho^2}{\xi^2}
\right)
\mu_B
B_z
\\
-
\dfrac{\xi}{2}
-
\dfrac{F^2\rho^2}{\xi}
+
\mathcal{O}(B^2_z,F^3)
\quad.
\end{multline}
The first term for the two limiting regimes contains the effective electronic g-factors given in Eqs.\eqref{eq.geff_weakSOC} and \eqref{eq.geff_strongSOC}, respectively. 

\subsubsection{First Excited State}
First-excited states of $\underline{\underline{H}}_\uparrow$ and $\underline{\underline{H}}_\downarrow$ (\textit{cf.} Eqs.\eqref{eq.rjt_hamilton_basis_up} and \eqref{eq.rjt_hamilton_basis_down}) correspond to $E^i_1(X)$ in Eq.\eqref{eq.transform_3x3_hamilton_l1} with
\begin{align}
E^i_1(X)
&=
E^p_1(X)
=
E^\uparrow_1
\quad,
\vspace{0.2cm}
\\
E^i_1(X)
&=
E^s_1(X)
=
E^\downarrow_1
\quad.
\end{align}
Here, we find
\begin{align}
E^\uparrow_1
&=
\dfrac{g_e
\mu_B}{2}
B_z
+
\sqrt{
\left(
\mu_B
B_z
\pm
\frac{\xi}{2}
\right)^2
+
F^2
\rho^2
}
\,,
\end{align}
and
\begin{align}
E^\downarrow_1
&=
-
\dfrac{g_e
\mu_B}{2}
B_z
+
\sqrt{
\left(
\mu_B
B_z
\mp
\frac{\xi}{2}
\right)^2
+
F^2
\rho^2
}
\,,
\end{align}
with sign convention as above. In the weak $B_z$-field regime, we have
\begin{multline}
E^\uparrow_1
=
\dfrac{1}{2}
\sqrt{
\xi^2
+
4
F^2
\rho^2
}
\\
+
\left(
\dfrac{g_e\mu_B}{2}
\pm
\dfrac{\mu_B\xi}{\sqrt{\xi^2+4F^2\rho^2}}
\right)
B_z
+
\mathcal{O}(B^2_z)
\quad,
\end{multline}
and
\begin{multline}
E^\downarrow_1
=
\dfrac{1}{2}
\sqrt{
\xi^2
+
4
F^2
\rho^2
}
\\
+
\left(
-
\dfrac{g_e\mu_B}{2}
\mp
\dfrac{\mu_B\xi}{\sqrt{\xi^2+4 F^2\rho^2}}
\right)
B_z
+
\mathcal{O}(B^2_z)
\quad,
\end{multline}
In weak $(\xi\ll F\rho)$ and strong $(\xi\gg F\rho)$ SOC regimes, we obtain for the $\uparrow$-sector
\begin{multline}
E^\uparrow_1
=
\dfrac{1}{2}
\left(
g_e
\pm
\dfrac{\xi}{F\rho}
\right)
\mu_B
B_z
\\
+
F\rho
+
\dfrac{\xi^2}{8F\rho}
+
\mathcal{O}(B^2_z,\xi^3)
\quad,
\end{multline}
and
\begin{multline}
E^\uparrow_1
=
\dfrac{1}{2}
\left(
g_e
\pm
2
\mp
\dfrac{4F^2\rho^2}{\xi^2}
\right)
\mu_B
B_z
\\
+
\dfrac{\xi}{2}
+
\dfrac{F^2\rho^2}{\xi}
-
\mathcal{O}(B^2_z,F^3)
\quad,
\end{multline}
while the $\downarrow$-sector is determined by
\begin{multline}
E^\downarrow_1
=
-
\dfrac{1}{2}
\left(
g_e
\pm
\dfrac{\xi}{F\rho}
\right)
\mu_B
B_z
\\
+
F\rho
+
\dfrac{\xi^2}{8F\rho}
+
\mathcal{O}(B^2_z,\xi^3)
\quad,
\end{multline}
and 
\begin{multline}
E^\downarrow_1
=
-
\dfrac{1}{2}
\left(
g_e
\pm
2
\mp
\dfrac{4F^2\rho^2}{\xi^2}
\right)
\mu_B
B_z
\\
+
\dfrac{\xi}{2}
+
\dfrac{F^2\rho^2}{\xi}
+
\mathcal{O}(B^2_z,F^3)
\quad.
\end{multline}

\setcounter{equation}{0}
\renewcommand{\theequation}{\thesubsection.\arabic{equation}}
\subsection{Details of the Effective Hamiltonian Formalism}
\label{sec.details_eff_hamilton}
Polariton and spectator Hamiltonians read in the truncated subspace approximation
\small{
\begin{widetext}
\begin{align}
\underline{\underline{\tilde{H}}}_p
=
\begin{pmatrix}
-
\dfrac{\xi}{2}
\mp
(1\mp\frac{g_e}{2})
\mu_B
B_z
& 
F\rho 
&
0
\vspace{0.2cm}
\\
F\rho 
& 
\dfrac{\xi}{2}
\pm
(1\pm\frac{g_e}{2})
\mu_B 
B_z
&
\dfrac{g_0\,g_e\mu_B}{2c}
\sqrt{\dfrac{\hbar\omega_c}{2}}
\vspace{0.2cm}
\\
0
&
\dfrac{g_0\,g_e\mu_B}{2c}
\sqrt{\dfrac{\hbar\omega_c}{2}}
&
-
\dfrac{\xi}{2} 
+
\hbar\omega_c
\pm
(1\mp\frac{g_e}{2})
\mu_B
B_z
\end{pmatrix}
\quad,
\label{eq.polariton_hamilton_3d}
\vspace{0.2cm}
\\
\underline{\underline{\tilde{H}}}_s
=
\begin{pmatrix}
-
\dfrac{\xi}{2}
\pm
(1\mp\frac{g_e}{2})
\mu_B
B_z
& 
F\rho 
&
0
\vspace{0.2cm}
\\
F\rho
& 
\dfrac{\xi}{2}
\mp
(1\pm\frac{g_e}{2})
\mu_B 
B_z
&
-
\dfrac{g_0\,g_e\mu_B}{2c}
\sqrt{\dfrac{\hbar\omega_c}{2}}
\vspace{0.2cm}
\\
0
&
-
\dfrac{g_0\,g_e\mu_B}{2c}
\sqrt{\dfrac{\hbar\omega_c}{2}}
&
-
\dfrac{\xi}{2} 
+
\hbar\omega_c
\mp
(1\mp\frac{g_e}{2})
\mu_B
B_z
\end{pmatrix}
\quad,
\label{eq.spectator_hamilton}
\end{align}
\end{widetext}
}\normalsize
where
\begin{align}
E^i_2(X)
&=
E^p_2(X)
=
-
\dfrac{\xi}{2} 
+
\hbar\omega_c
\pm
(1\mp\frac{g_e}{2})
\mu_B
B_z
\,,
\vspace{0.2cm}
\\
E^i_2(X)
&=
E^s_2(X)
=
-
\dfrac{\xi}{2} 
+
\hbar\omega_c
\mp
(1\mp\frac{g_e}{2})
\mu_B
B_z
\,.
\end{align}
Rotation matrix elements take in the weak B-field limit the form
\begin{align}
\sin\frac{\varphi_p([3e^1])}{2}
&=
-
\sin\varphi^\star
+
\dfrac{2 F\rho\cos\varphi^\star}{\xi^2+4F^2\rho^2}
\mu_B
B_z
+
\mathcal{O}(B^2_z)
\quad,
\label{eq.sinp_weak_B}
\vspace{0.2cm}
\\
\sin\frac{\varphi_s([3e^1])}{2}
&=
-
\sin\varphi^\star
-
\dfrac{2 F\rho\cos\varphi^\star}{\xi^2+4F^2\rho^2}
\mu_B
B_z
+
\mathcal{O}(B^2_z)
\quad,
\label{eq.sins_weak_B}
\end{align}
with
\begin{align}
\varphi^\star
&=
\arctan\left(\dfrac{2F\rho}{\xi}\right)
\quad.
\end{align}
We consider now Taylor expansion around $\xi=0$ and $F=0$ to obtain approximations for weak $(\xi\ll F\rho)$ and strong $(\xi\gg F\rho)$ SOC regimes given by
\begin{align}
\sin\frac{\varphi_p([3e^1])}{2}
&=
-
\dfrac{1}{\sqrt{2}}
+
\dfrac{1}{\sqrt{2}}
\dfrac{\mu_BB_z}{2F\rho}
+
\mathcal{O}(B^2_z,\xi^2)
\,,
\label{eq.sinp_weak_B_weakSOC}
\vspace{0.2cm}
\\
\sin\frac{\varphi_p([3e^1])}{2}
&=
-
\dfrac{F\rho}{\xi}
+
\dfrac{2 F\rho}{\xi^2}
\mu_B
B_z
+
\mathcal{O}(B^2_z,F^2)
\,,
\label{eq.sinp_weak_B_strongSOC}
\end{align}
and 
\begin{align}
\sin\frac{\varphi_s([3e^1])}{2}
&=
-
\dfrac{1}{\sqrt{2}}
-
\dfrac{1}{\sqrt{2}}
\dfrac{\mu_BB_z}{2F\rho}
+
\mathcal{O}(B^2_z,\xi^2)
\,,
\label{eq.sins_weak_B_weakSOC}
\vspace{0.2cm}
\\
\sin\frac{\varphi_s([3e^1])}{2}
&=
-
\dfrac{F\rho}{\xi}
-
\dfrac{2 F\rho}{\xi^2}
\mu_B
B_z
+
\mathcal{O}(B^2_z,F^2)
\,.
\label{eq.sins_weak_B_strongSOC}
\end{align}
From Eq.\eqref{eq.rotation_angles}, we find
\begin{align}
\sin\frac{\varphi_p([3e^1])}{2}
&=
\sin\frac{\varphi_s([2e^3])}{2}
\quad,
\vspace{0.2cm}
\\
\sin\frac{\varphi_s([3e^1])}{2}
&=
\sin\frac{\varphi_p([2e^3])}{2}
\quad.
\end{align}
We obtain the excitation energy, $\Delta_{02}$, in second-order QDPT from zero-order energies of polariton and spectator problems as
\begin{align}
\begin{matrix}
\Delta^p_{02}
&=
\hbar\omega
\pm
(1\mp\frac{g_e}{2})
\mu_B
B_z
\approx
\hbar\omega
\vspace{0.2cm}
\\
\Delta^s_{02}
&=
\hbar\omega
\mp
(1\mp\frac{g_e}{2})
\mu_B
B_z
\approx
\hbar\omega
\end{matrix}
\quad,
\end{align}
for $\hbar\omega_c>\mu_BB_z$ such that $\Delta^p_{02}\approx\Delta^s_{02}=\Delta_{02}$.

\setcounter{equation}{0}
\renewcommand{\theequation}{\thesubsection.\arabic{equation}}
\subsection{Dimensional Analysis}
\label{sec.dim_anal}
We show that 
\begin{align}
\left[
\dfrac{g^2_0 g^2_e\mu^2_B}{c^2}
\right]
=
E_h
\quad.
\end{align}
To this end, we exploit
\begin{align}
[g_0]
=
\dfrac{\sqrt{E_h}}{e a_0}
\quad,\,
[\mu_B]
=
\dfrac{e\hbar}{m_e}
\quad,\,
[c]
=
\dfrac{a_0 E_h}{\hbar}
\quad,
\end{align}
and $[g_e]=-$ which allows us to write
\begin{align}
\left[
\dfrac{g^2_0 g^2_e\mu^2_B}{c^2}
\right]
&=
\dfrac{E_h}{e^2 a^2_0}
\dfrac{e^2\hbar^2}{m^2_e}
\dfrac{\hbar^2}{a^2_0 E^2_h}
=
\dfrac{\hbar^4}{E_h a^4_0 m^2_e}
\,,
\end{align}
such that with $[E_h]=\frac{\hbar^2}{m_e a^2_0}$, the desired result follows as
\begin{align}
\left[
\dfrac{g^2_0 g^2_e\mu^2_B}{c^2}
\right]
&=
\dfrac{E^2_h}{E_h}
=
E_h
\quad.
\end{align}



\end{document}